\documentclass[letterpaper, 10 pt, conference]{ieeeconf}  

\IEEEoverridecommandlockouts                              

\usepackage{cite}
\usepackage{amsmath,amssymb,amsfonts}
\usepackage{algorithmic}
\usepackage{graphicx}
\usepackage{textcomp}
\usepackage{booktabs}
\usepackage{multirow}
\usepackage{xcolor}

\usepackage{amsthm}
\usepackage{thmtools}
\usepackage[ruled,vlined]{algorithm2e}
\usepackage{color}
\usepackage{hyperref}

\usepackage{dblfloatfix}

\title{\LARGE \bf
A Dynamic Parametric Simulator for Fetal Heart Sounds
}

\author{Yingtong Zhou$^{1}$, Yiang Zhou$^{1}$, Zhengxian Qu$^{1}$, Kang Liu$^{1}$ and Ting Tan$^{1}$
\thanks{*This work is supported by Aerie Intelligent Technology Corp, USA}
\thanks{$^{1}$ Yingtong Zhou, Yiang Zhou are intern students at Aerie Intelligent Technology Corp, USA. Zhengxian Qu, Kang Liu and Ting Tan are data scientists at Aerie Intelligent Technology Corp, USA. Corresponding email: {\tt\small tting@aerieintech.com}.}
}

\begin{document}
\maketitle
\thispagestyle{empty}
\pagestyle{empty}

\begin{abstract}
Research on fetal phonocardiogram (fPCG) is challenged by the limited number of abdominal recordings, substantial maternal interference, and marked transmission-induced signal attenuation that complicate reproducible benchmarking. We present a reproducible dynamic parametric simulator that generates long abdominal fPCG sequences by combining cycle-level fetal S1/S2 event synthesis with a convolutional transmission module and configurable interference and background noise. Model parameters are calibrated cycle-wise from real abdominal recordings to capture beat-to-beat variability and to define data-driven admissible ranges for controllable synthesis. The generated signals are validated against real recordings in terms of envelope-based temporal structure and frequency-domain characteristics. The simulator is released as open software to support rapid, reproducible evaluation of fPCG processing methods under controlled acquisition conditions.


\end{abstract}


\section{Introduction}

The fetal phonocardiography (fPCG), recordings of fetal heartbeat sounds from the maternal abdomen, provides a non-invasive, passive modality for fetal cardiac monitoring involving no transmitted energy to the fetus. Its low-cost and longitudinal acquisition with low risk motivates data-driven methods for fetal heart rate estimation and broader signal analysis \cite{muller2025standardized,Kovacs2005LongTerm,adithya2017trends}. Moreover, fPCG signals carry rich physiological information: in addition to yielding the fetal heart rate (FHR), they contain the detailed heart sound waveforms and can reveal phenomena like murmurs or split heart sounds that are invisible to ultrasonographic cardiotocography (CTG). For these reasons, there is strong motivation to incorporate fPCG into monitoring of fetal health\cite{Barnova2022Comparative}.

In practice, abdominal fPCG is not yet a routine clinical tool, primarily due to a poor signal-to-noise ratio (SNR) and strongly affected by maternal cardiac activity and other nonstationary artifacts, making robust signal processing and reliable clinical interpretation challenging \cite{CESARELLI2012513,Chourasia2009DevelopmentOA,Gerhardt2000FetalExposures}.
Moreover, publicly available fPCG datasets remain limited in size and recording diversity \cite{CESARELLI2012513,Samieinasab2015BSS,Bhaskaran2022HR}, constraining reproducible benchmarking. To sidestep these limitations, controllable simulators that can generate clinically plausible signals under specified interference and SNR conditions have been emerging\cite{moya2025challenges}. However, existing fPCG simulators are useful for controlled testing, yet they are typically constructed without explicit modeling of abdominal sound transmission or quantitative alignment with real abdominal recordings, which limits the authenticity and reliability of simulated signals
\cite{Ruffo2010ITAB,ruffo2010algorithm,Romano2015Tool,Amrutha2019FoetalAcousticSimulator}.

In this study, we present a reproducible dynamic parametric simulator for abdominal fPCG that couples event-level cardiac sound synthesis with an explicit abdomen-to-sensor transmission module. 
The simulator uses interpretable parameters with variability calibrated from real abdominal recordings, allowing controlled long-sequence synthesis while preserving beat-to-beat variation. Validation is performed by applying the same preprocessing pipeline to both the simulated signals and real fPCG recordings, using envelope-based temporal statistics (cycle-averaged autocorrelation function, ACF) and frequency-domain consistency (normalized power spectral density, PSD). The implementation is released as both a Python package and an interactive web front end.

\section{Method}
\label{sec:method}

\subsection{Overview}
We present a dynamic parametric simulator for abdominal fPCG signals. The simulator is designed around an explicit generative decomposition, interpretable event-level parametrization of fetal and maternal heart sounds, a convolutional propagation model for abdominal transmission, and a controllable noise mechanism. The pipeline supports two usage modes: (i) preset-driven interactive synthesis for rapid scenario exploration and (ii) configuration-driven batch generation for research workflows. Empirical calibration and quantitative validation are described in Sec.~\ref{sec:experiments}.

\subsection{Signal model}
An observed abdominal fPCG signal is modeled, at the first-order approximation, as a superposition of fetal heart sounds, maternal cardiac interference, and residual noise.
\begin{equation}
x(t) = h(t)*(x_f(t) + x_m(t)) + n(t).
\label{eq:mix}
\end{equation}
Here, $x_f(t)$ represents the fetal heart sound (FHS), dominated by the first heart sound (S1) and second heart sounds (S2), which are standard PCG events associated with the closure of the mitral/tricuspid valves\cite{Kahankova2023Review}; $x_m(t)$ represents maternal heart sounds; $n(t)$ aggregates non-cardiac physiological artifacts and measurement noise; and $h(t)$ denotes the overall abdominal transmission (and sensor) impulse response (implemented in Sec.~\ref{sec:prop} as a cascaded convolutional model). Our simulator generates $(x_f, x_m, n)$ separately and combines them to form the output $x(t)$.

\subsection{Event-based parametrization of fetal heart sounds}
\subsubsection{Cycle timing and event placement}
Within each beat, let $\{t_k\}_{k\ge 1}$ denote fetal beat onset times, and $\Delta T^{(k)}$ denote the systolic interval during S1 and S2 in beat $k$.
\begin{equation}
x_f(t) = \sum_{k}\Big[
\phi\!\left(t-t_k;\,\theta_{\mathrm{S1}}^{(k)}\right)
+
\phi\!\left(t-t_k-\Delta T^{(k)};\,\theta_{\mathrm{S2}}^{(k)}\right)
\Big],
\label{eq:fetal_sum}
\end{equation}
where $\theta_{\mathrm{S1}}^{(k)}$ and $\theta_{\mathrm{S2}}^{(k)}$ are event parameters for cycle $k$. The onset sequence $\{t_k\}$ can be specified by a constant FHR, by a user-defined beat-to-beat interval (RR) series, or by a weak heart-rate-variability (HRV) process described in Sec.~\ref{sec:hrv}.

\subsubsection{Asymmetric damped-sinusoid kernel}
Each heart-sound event is generated using a compact asymmetric kernel
as a parametric representation with a short attack and exponential decay,
\begin{equation}
\phi(t;A,f_0,T_a,\tau)=A\sin(2\pi f_0 t)\,a(t;T_a,\tau),
\label{eq:kernel}
\end{equation}
with envelope
\begin{equation}
a(t;T_a,\tau)=
\begin{cases}
\frac{t}{T_a}, & 0\le t<T_a,\\[2pt]
\exp\!\left(-\frac{t-T_a}{\tau}\right), & t\ge T_a,\\[2pt]
0, & t<0.
\end{cases}
\label{eq:env_kernel}
\end{equation}
Here $A$ is amplitude, $f_0$ is carrier frequency, $\tau$ controls decay and effective duration, the attack time $T_a$ is fixed to a constant (default $T_a=8$~ms) to reduce parameter coupling with $\tau$ and improve identifiability while preserving the rapid onset of S1/S2 events.
The use of a damped-sinusoid kernel is motivated by the characteristic morphology of heart sounds observed in phonocardiogram recordings, where each S1 or S2 event typically appears as a short burst of quasi-periodic oscillations followed by rapid attenuation. Such transient oscillatory behavior has been reported in clinical and signal-processing studies of cardiac auscultation signals and is commonly modeled using exponentially damped sinusoids or closely related kernels \cite{Ali2011Dynamic,Choklati2022LaplacePCG}. 
This kernel family provides a low-dimensional parameterization that is straightforward to manipulate for normal and abnormal scenarios. 

\subsubsection{Per-cycle parameter vector}
The kernel-parameter vectors in Eq.~(2) are defined as
\begin{equation}
\begin{aligned}
\boldsymbol{\theta}_{\mathrm{S1}}^{(k)} &:= \left[A_{\mathrm{S1}}^{(k)},\, f_{0,\mathrm{S1}},\, T_a,\, \tau_{\mathrm{S1}}^{(k)}\right],\\
\boldsymbol{\theta}_{\mathrm{S2}}^{(k)} &:= \left[A_{\mathrm{S2}}^{(k)},\, f_{0,\mathrm{S2}},\, T_a,\, \tau_{\mathrm{S2}}^{(k)}\right].
\end{aligned}
\end{equation}
where \(f_{0,\mathrm{S1/S2}}\) and \(T_a\) are preset hyperparameters shared across cycles. Accordingly, the cycle-wise adjustable parameters are collected in
\begin{equation}
\boldsymbol{\theta}^{(k)}=
\left[
A_{\mathrm{S1}}^{(k)},\,A_{\mathrm{S2}}^{(k)},\,
\tau_{\mathrm{S1}}^{(k)},\,\tau_{\mathrm{S2}}^{(k)},\,
\Delta T^{(k)}
\right].
\label{eq:theta}
\end{equation}

In the default synthesis mode, the carrier frequency $f_0$ is treated as a preset hyperparameter for fetal S1/S2, while $(A,\tau,\Delta T)$ vary across cycles through the sampling mechanism in Sec.~\ref{sec:sampling}. To promote stable envelope morphology for downstream envelope-based diagnostics, we optionally use a shared envelope for S1 and S2 during synthesis,
\begin{equation}
\tau_{\mathrm{common}}^{(k)}=\frac{\tau_{\mathrm{S1}}^{(k)}+\tau_{\mathrm{S2}}^{(k)}}{2},
\label{eq:sharedtau}
\end{equation}
and generate S1/S2 using $\tau_{\mathrm{common}}^{(k)}$ while keeping the amplitude ratio free. This option is used only at generation time and does not change the definition of $\boldsymbol{\theta}^{(k)}$.

\subsection{Maternal heart sound component}
Maternal heart sounds are often stronger than fetal heart sounds and constitute the dominant endogenous interference in abdominal fPCG recordings \cite{Kovacs2011Review}.
Maternal cardiac interference is modeled as a second two-event source process,
\begin{equation}
x_m(t)=\sum_{\ell}\Big[
\phi\!\left(t-\tilde t_\ell;\,\tilde\theta_{\mathrm{S1}}^{(\ell)}\right)
+
\phi\!\left(t-\tilde t_\ell-\Delta \tilde T^{(\ell)};\,\tilde\theta_{\mathrm{S2}}^{(\ell)}\right)
\Big],
\label{eq:maternal_sum}
\end{equation}
where $\{\tilde t_\ell\}$ denote maternal beat times determined by the maternal heart rate (MHR). 
In the current single-channel setting, maternal parameters $\tilde\theta$ are treated as fixed nuisance hyperparameters, with only a global scaling factor exposed to control interference strength. 
This design preserves a compact and interpretable fetal parameterization without introducing additional fitted degrees of freedom.

\subsection{Abdominal propagation model}
\label{sec:prop}
\subsubsection{Cascaded impulse response}
We approximate abdominal transmission by a causal linear time-invariant model applied to cardiac sources. The received cardiac component is
\begin{equation}
x_c(t)= \left(h*(x_f+x_m)\right) (t),
\label{eq:conv_all}
\end{equation}
where $*$ denotes convolution and $h(t)$ is the effective impulse response. We model $h(t)$ as a cascade of two exponentially decaying kernels,
\begin{equation}
h_1(t)=A_1 e^{-\beta_1 t}\,\mathbb{I}(t\ge 0),\quad
h_2(t)=A_2 e^{-\beta_2 t}\,\mathbb{I}(t\ge 0),
\label{eq:h12}
\end{equation}
and
\begin{equation}
h(t)=\frac{(h_1*h_2)(t)}{\int_{0}^{\infty}(h_1*h_2)(u)\,du}.
\label{eq:htotal}
\end{equation}
The normalization fixes overall gain, leaving attenuation and smoothing effects determined by $(A_1,\beta_1,A_2,\beta_2)$ up to scale. Using a shared $h(t)$ for fetal and maternal sources avoids inconsistent spectral shaping across components. Note that here, the abdominal sound transmission is modeled as linear and time-invariant (LTI), assuming local quasi-stationarity of fetal heart sound morphology within a short temporal window. 

\subsection{Noise and artifact model}
Residual disturbances are modeled using a simplified colored stochastic process with optional slow gain modulation. This noise model is intended to capture the dominant characteristics observed in the considered abdominal phonocardiogram recordings, rather than representing all possible physiological or environmental noise sources.
We first generate AR(1) (first-order autoregressive) noise
\begin{equation}
n_t = \rho n_{t-1} + \sqrt{1-\rho^2}\,\varepsilon_t,\qquad
\varepsilon_t\sim\mathcal{N}(0,1),\quad |\rho|<1,
\label{eq:ar1}
\end{equation}
and optionally apply a low-frequency gain envelope
\begin{equation}
g(t)=1+\gamma\,\mathrm{LP}\{w(t)\},\qquad \tilde n(t)=g(t)\,n(t),
\label{eq:gain}
\end{equation}
where $w(t)$ is white noise and $\mathrm{LP}\{\cdot\}$ is a low-pass operator. The final simulated signal is
\begin{equation}
x(t)=x_c(t)+\sigma_n \tilde n(t).
\label{eq:finalsig}
\end{equation}
When a target SNR (in dB) is specified, $\sigma_n$ is set using RMS scaling,
\begin{equation}
\sigma_n=\frac{\mathrm{RMS}(x_c)}{10^{\mathrm{SNR}/20}}.
\label{eq:snr}
\end{equation}
This formulation provides a controlled mechanism for adjusting noise intensity. More detailed and source-specific noise models are left for future work.

\subsection{Sampling of cycle-to-cycle variability}
\label{sec:sampling}
To introduce beat-to-beat variability, the simulator samples one cycle-level fetal parameter vector $\boldsymbol{\theta}^{(k)}$ for each cardiac cycle from user-specified admissible ranges $[\mathbf{l}_s,\mathbf{u}_s]$,
\begin{equation}
\boldsymbol{\theta}^{(k)} \sim p(\boldsymbol{\theta}\mid \mathbf{l}_s,\mathbf{u}_s),
\end{equation}
where $p(\cdot)$ is a user-selected prior within the bounds. 
We support a flat (uniform) prior for broad coverage and a truncated Gaussian prior to concentrate samples around typical values. At the cycle level, sampled parameters control the S1/S2 event generation, while global parameters govern propagation and noise; all components are then combined to form the simulated fPCG signal.

\subsection{Heart-rate specification and weak HRV}
\label{sec:hrv}
Cycle onset times are determined by an RR series. In the simplest case, RR is constant, $RR_k=\overline{RR}$. To introduce weak HRV while keeping the simulator low-dimensional, we optionally use
\begin{equation}
RR_k=\overline{RR}+\alpha d_k+\eta_k,
\label{eq:hrv}
\end{equation}
where $d_k$ is a low-frequency drift process obtained by smoothing i.i.d. noise, $\eta_k$ is small jitter, and $\alpha$ controls the variability scale. Each synthesized cycle can be time-stretched to match $RR_k$ prior to concatenation. The RR series determines the cycle onset times $\{t_k\}$ via
\begin{equation}
t_{k+1}=t_k+RR_k,
\end{equation}
(i.e., $t_k=\sum_{i=1}^{k-1} RR_i$).
Each synthesized cycle can be time-stretched to match $RR_k$ prior to concatenation.

\subsection{Scenario control and abnormal presets}
Abnormal scenarios are generated by applying controlled perturbations to selected parameters while keeping the remaining configuration fixed. Examples include modifying $\Delta T$ to alter S1 -- S2 systolic interval, changing the amplitude ratio $A_{\mathrm{S2}}/A_{\mathrm{S1}}$, or adjusting propagation and SNR settings to emulate more challenging abdominal conditions. All scenario definitions are implemented as parameter presets and are therefore reproducible.

\subsection{Software packaging and deployment}
\begin{figure*}[t]
    \centering
    \includegraphics[width=0.8\linewidth]{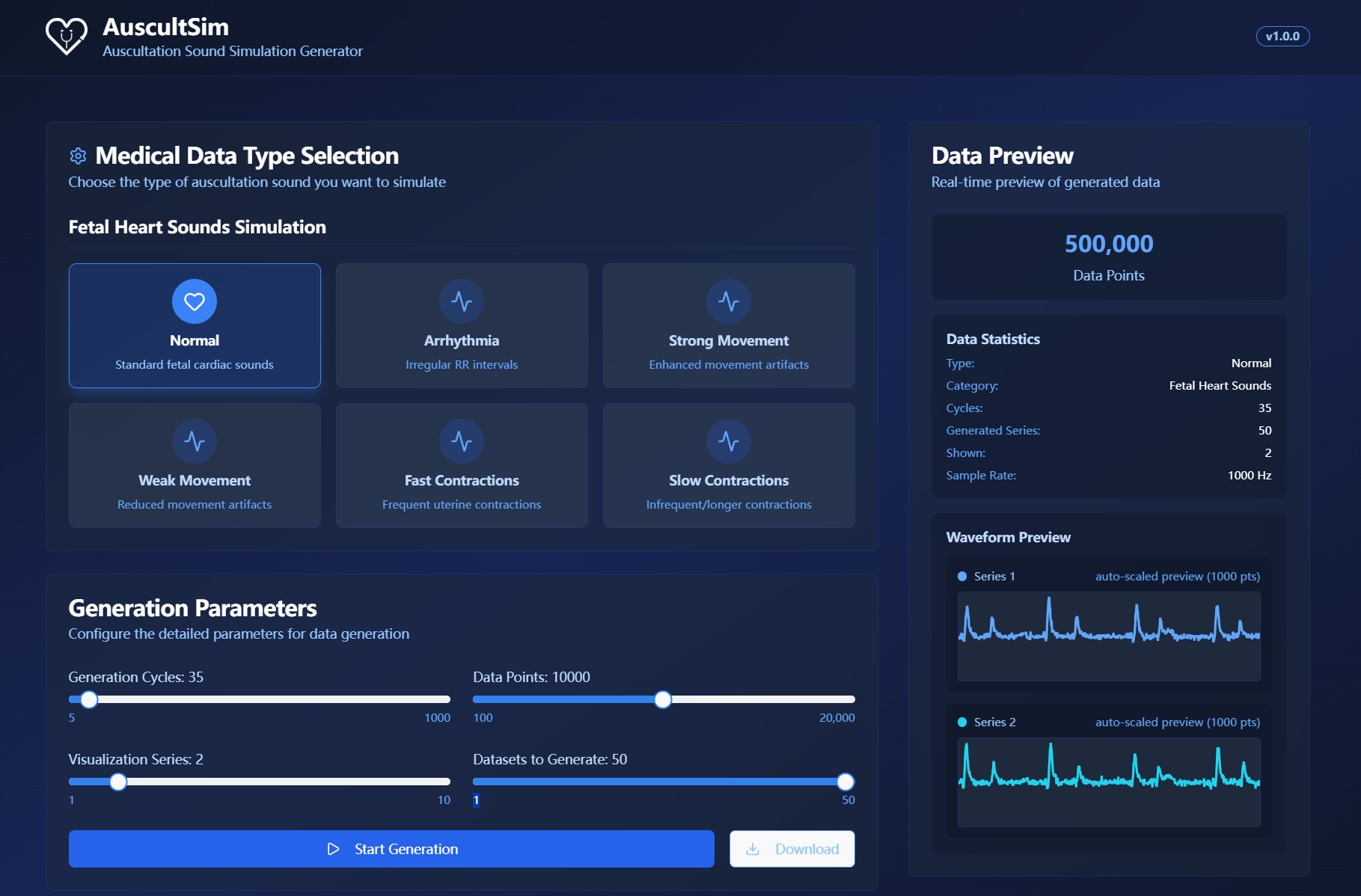}
    \caption{Interactive web front end of preset-based fPCG generation.}
    \label{fig:frontend}
\end{figure*}
The simulator is provided as (i) a configuration-driven Python package and (ii) a preset-based interactive web front end for rapid synthesis and visualization \footnote{\url{https://ytzhou123.github.io/FH-Simulator/}}, present in Figure~\ref{fig:frontend}. The package reads parameters from an INI configuration file and exposes functions for generating fetal and maternal sources, applying propagation, adding noise, and exporting waveforms. The web front end uses a fixed set of vetted presets and provides interactive controls for sequence length, sampling mode, and noise level. The calibration procedure and quantitative evaluation protocol are reported in Sec.~\ref{sec:experiments}.


\section{Validation experiments}
\label{sec:experiments}

\subsection{Dataset and Preprocessing Setup}
Validation experiments were conducted on the \emph{Fetal PCG Database} released on PhysioNet (26 abdominal fPCG recordings, gestational week 31--40; sampling rate 333~Hz; normalized device output) \cite{CESARELLI2012513,goldberger2000physiobank}. We report results on a representative record, selected for its relatively stable cardiac activity.

For each selected segment, we applied a bandpass filter (20--150~Hz) and computed the Hilbert envelope, followed by low-pass smoothing at 8~Hz and normalization. This envelope-level smoothing suppresses carrier-frequency fluctuations while preserving the dominant cardiac cycle structure. The dominant cardiac period $T_0$ was estimated from the envelope autocorrelation within a physiologically plausible FHR range (80--200~bpm). Cycle boundaries were obtained using envelope-peak detection with minimum-distance constraints relative to $T_0$, followed by onset refinement via a local threshold rule, discarding implausible RR intervals (0.25--0.90~s). The resulting cycles were zero-mean normalized and used for parameter fitting; multi-cycle signals for validation were formed by concatenating retained cycles from the selected stable window.

\subsection{Fitting and Parameter Sampling}

Model parameters were estimated from real abdominal fPCG segments on a per-cycle basis. For each extracted cardiac cycle, the fetal S1/S2 parameters $\boldsymbol{\theta}^{(k)}=(A_{S1},A_{S2},\tau_{S1},\tau_{S2},\Delta T)$ were fitted independently for each extracted cycle. Stochastic noise components were disabled during fitting to preserve parameter identifiability. The fitted parameter set $\{\boldsymbol{\theta}^{(k)}\}_{k=1}^{K}$ was summarized to define a data-driven sampling region. 
Cycle-wise parameters were estimated by nonlinear least squares using SciPy ~\cite{Virtanen2020SciPy}. 
To visualize inter-parameter correlations, we fitted a multivariate Gaussian to the fitted set $\{\boldsymbol{\theta}^{(k)}\}$ using the sample covariance and drew Monte Carlo samples to produced marginal densities and pairwise contours. 
For multi-cycle synthesis and validation, we constructed a data-driven sampling region around the best-fit parameters by taking a bounded box defined from the empirical dispersion of $\{\boldsymbol{\theta}^{(k)}\}$ (clipped to global physiologic bounds). Parameters were then sampled per cycle using either bounded uniform sampling or, optionally, an affine-invariant ensemble MCMC sampler (\texttt{emcee}) targeting the same bounded uniform distribution~\cite{GoodmanWeare2010Ensemble,ForemanMackey2013emcee}.
S1 -- S2 systolic interval was optionally stabilized via bootstrap sampling from envelope peak-to-peak measurements of real cycles. For long-sequence synthesis, per-cycle signals were time-stretched to match a target RR series (either the empirical RR sequence or a controlled HRV variant

\subsection{Evaluation}
To validate the realism of the proposed simulator, we applied the same preprocessing step for real and simulated signals. We compared (i) a multi-cycle envelope waveform, where both real and simulated signals are converted to a normalized Hilbert envelope after smoothing, 
(ii) the cycle-averaged envelope autocorrelation, computed by extracting envelopes per cycle, forming a normalized ACF for each cycle, and averaging after zero-truncation to a common length, and (iii) the normalized PSD estimated by \cite{welch2003use} after bandpass filtering and RMS normalization. 
In addition, parameter variability and inter-parameter dependence were summarized using a corner plot built from the fitted parameter distribution, with the sampled parameter vectors overlaid to verify that generation remains within the calibrated sampling bounds.

\section{Results}

\begin{figure*}[t]
    \centering
    \includegraphics[width=0.8\linewidth]{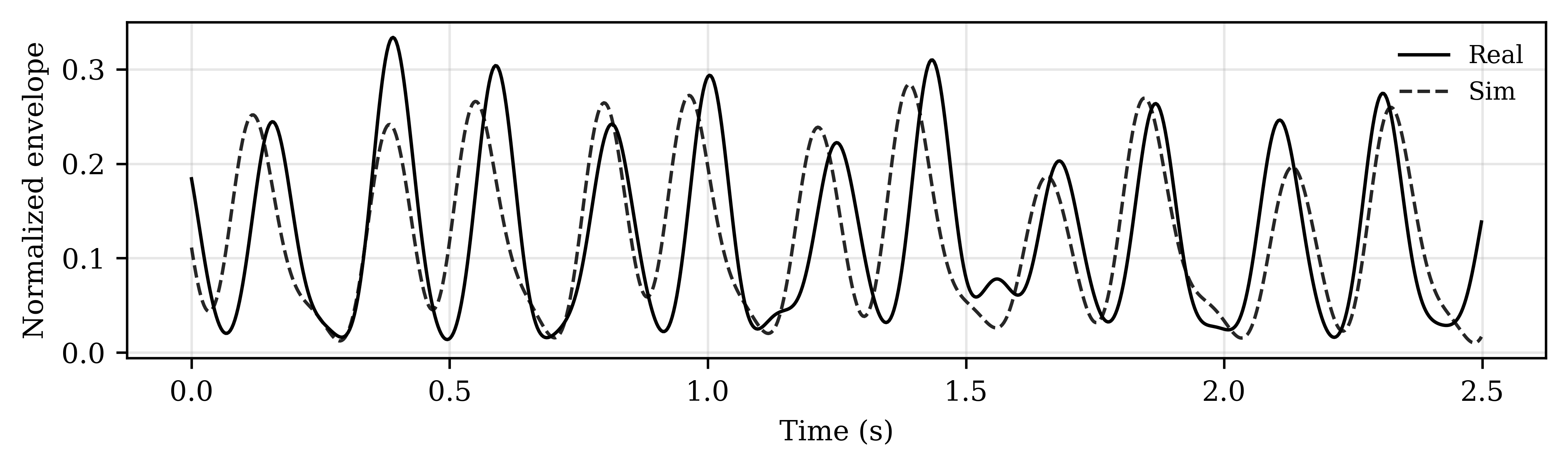}
    \caption{Envelope comparison between real and simulated multi-cycle signals on a representative stable window.}
    \label{fig:res_env}
\end{figure*}

\begin{figure*}[t]
    \centering
    \includegraphics[width=0.47\linewidth]{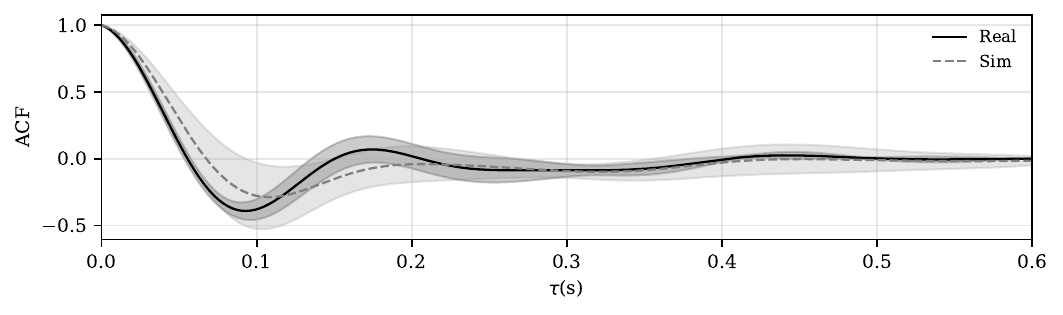}\hfill
    \includegraphics[width=0.47\linewidth]{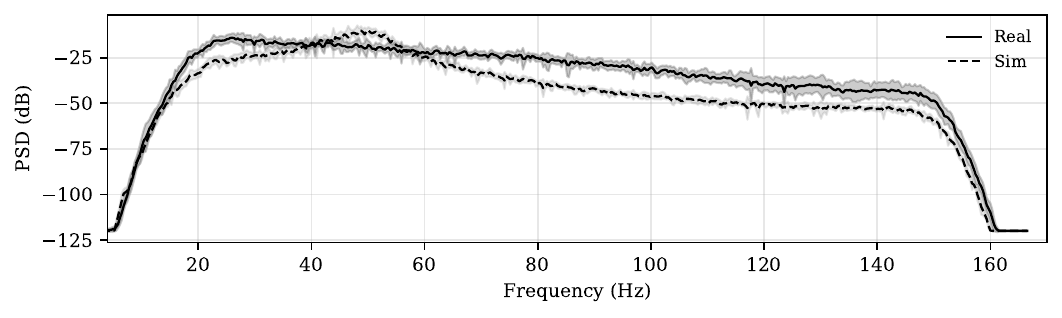}
    \caption{Comparison between real and simulated signals under identical preprocessing:
    (left) cycle-averaged envelope ACF;
    (right) PSD (dB) with error band.}
    \label{fig:res_acf_psd}
\end{figure*}

\begin{figure*}[t]
    \centering
    \includegraphics[width=0.8\linewidth]{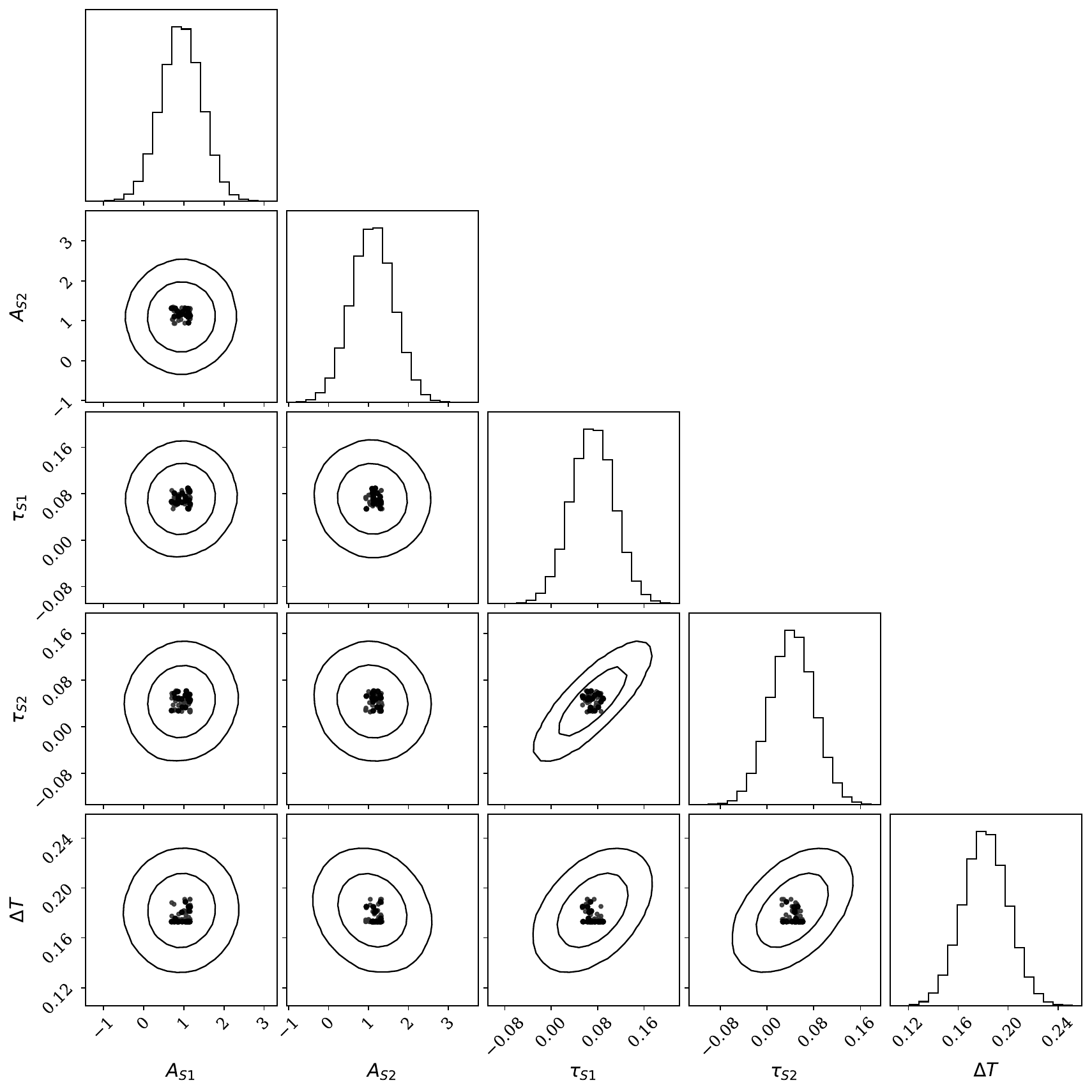}
    \caption{Corner plot of fitted parameters $\{A_{S1},A_{S2},\tau_{S1},\tau_{S2},\Delta T\}$.}
    \label{fig:res_corner}
\end{figure*}

\subsection{Multi-cycle Fitting Consistency}
\label{subsec:res_multicycle}
Fig.~\ref{fig:res_env} reports the envelope-level agreement between the real fPCG segment and the simulated multi-cycle signal generated from fitted and sampled parameters. Across the displayed window, the simulator reproduces the dominant periodic pattern and the beat-to-beat amplitude modulation, indicating that the proposed cycle segmentation and parameterized S1/S2 event model preserve the macro temporal structure of the recording. In particular, the timing of successive envelope peaks is largely consistent, suggesting that the fitted RR structure and the S1--S2 systolic interval constraints within each cycle are effectively maintained in long-sequence synthesis. Residual discrepancies are mainly observed as small peak misalignments and local amplitude differences, which can be attributed to imperfect cycle boundary detection and simplified propagation/noise assumptions.

\subsection{Two point statistics: ACF and PSD}
To quantify temporal structure beyond visual inspection, Fig.~\ref{fig:res_acf_psd} compares the cycle-averaged ACF and PSD under identical preprocessing. 
Both ACF and PSD were computed and summarized across segments.
The ACF curves exhibit similar early-lag behavior, including the negative trough and subsequent rebound, indicating that the simulator captures the characteristic intra-cycle spacing and envelope smoothness after averaging across cycles. Minor deviations in the secondary lobe magnitude suggest remaining differences in cycle-to-cycle variability, which is expected given the bounded sampling strategy and residual nonstationary interference in the real recordings.

The PSD comparison shows that the simulated signal preserves the overall spectral profile within the analysis band, with energy concentrated in the low-to-mid frequency range typical for abdominal fPCG. 
To avoid overly inflated error bars on a linear scale due to the low-frequency noise floor, PSD statistics were computed in the dB domain.
The simulated spectrum slightly underestimates power in part of the mid/high band, consistent with using compact source kernels and a coarse cascaded attenuation model. Nevertheless, the global roll-off and passband behavior remain consistent, supporting that the simulator produces signals with realistic frequency-domain characteristics suitable for algorithm development and controlled benchmarking.

\subsection{Parameter Distribution and Correlation Structure}
Fig.~\ref{fig:res_corner} visualizes the fitted parameter distribution and inter-parameter dependencies that motivate the preset ranges used for sampling and front-end presets. The marginal histograms are unimodal, indicating stable estimation under the robust fitting protocol. The joint contours reveal interpretable relationships: the duration parameters ($\tau_{S1}$ and $\tau_{S2}$) show a clear positive correlation, consistent with shared envelope smoothness constraints, while the amplitude parameters ($A_{S1}$ and $A_{S2}$) exhibit compensation effects reflecting trade-offs in matching waveform energy. $\Delta T$ presents comparatively weaker coupling with the remaining parameters in this dataset segment, supporting its separate stabilization (e.g., envelope peak-distance bootstrap) during long-sequence synthesis. Overall, these statistics provide a data-driven basis for calibrated sampling intervals and explain how the default preset configurations are derived from real recordings.

\section{Discussion}

The proposed simulator aims to provide a controllable and reproducible way to generate abdominal fPCG signals for algorithm testing. While the model reproduces key temporal and spectral statistics on a representative PhysioNet record under matched preprocessing, several limitations remain.

First, the maternal component and residual disturbances are represented in a simplified manner. In single-channel recordings, maternal activity and other abdominal artifacts can be broadband and time-varying. Our current formulation captures interference strength and basic spectral coloring, but does not cover the full variability observed in practice. Extending the interference model using additional recordings, and potentially multi-channel data when available, would improve realism.

Second, abdominal transmission is approximated by a causal LTI convolutional module. This is a pragmatic assumption for short analysis windows, but true transmission can vary with posture, sensor placement, and fetal movement. Future work could consider time-varying propagation or condition-specific transmission parameters, provided that sufficient data are available to constrain these extensions.

Third, calibration and validation were performed on a limited subset of recordings. Public datasets are often collected in late gestation and under specific acquisition setups, and model parameters may not directly transfer across devices or cohorts. Broader evaluation across records and datasets would be needed to quantify generalization and to derive more robust parameter ranges.

Despite these limitations, the simulator provides an interpretable parametric framework with data-driven calibration and open implementation. We expect it to be useful for controlled benchmarking of fPCG processing methods under specified interference and SNR conditions.

\section{Conclusion}
This work presented a reproducible dynamic parametric simulator for fetal phonocardiogram (fPCG) signals. The simulator generates fetal S1/S2 events with interpretable parameters, superposes a maternal cardiac component, and models transmission using a causal convolutional module, with configurable noise to emulate practical recording conditions. Cycle-level calibration on PhysioNet abdominal recordings was used to derive data-driven admissible parameter ranges and to support long-sequence synthesis with realistic variability. Under identical preprocessing, the simulated signals match the real recordings in envelope-level temporal structure and frequency-domain characteristics, as reflected by agreement in cycle-averaged ACF, and PSD. The simulator is released as a Python package and an interactive web front end to facilitate practical use, parameter exploration, and reproducible benchmarking
of normal and abnormal fPCG processing methods.

\section*{Acknowledgment}
This work is supported by Aerie Intelligent Technology Corp, USA.





\section*{APPENDIX A}

The simulator exposes a set of user-configurable parameters that control fetal cardiac events, abdominal transmission, maternal interference, and additive noise. These parameters are specified through a configuration interface and are used consistently across batch generation and interactive synthesis. Table~\ref{tab:user_params_default} lists the configurable parameters together with their default values and suggested ranges. The default settings correspond to physiologically plausible conditions, while the recommended ranges are informed by prior studies on fetal heart sounds and related physiological activity \cite{von2013normal,ramadan2020characteristics,crellin2025sonography,alalaf2022intrapartum,lai2016fetal,liu2010uterine}.

\begin{table}[htbp]
\centering
\renewcommand{\arraystretch}{1.2}
\begin{tabular}{l @{\hspace{8pt}} l @{\hspace{8pt}} l}
\toprule
\textbf{Parameter} & \textbf{Default} & \textbf{Suggested Range} \\
\midrule

\multicolumn{3}{l}{\textit{Dataset / Sampling}} \\
\midrule
\texttt{num\_samples} & 10 & -- \\
\texttt{cycles\_per\_sample} & 100 & -- \\
\texttt{fs} & 1000 Hz & 500--2000 Hz \\

\midrule
\multicolumn{3}{l}{\textit{Fetal / Maternal Heart Rate}} \\
\midrule
\texttt{fhr} & 140 bpm & 120--160 bpm \\
\texttt{mhr} & 80 bpm & 60--100 bpm \\

\midrule
\multicolumn{3}{l}{\textit{Transmission Model}} \\
\midrule
\texttt{r1} & 0.01 m & 0.005--0.02 m \\
\texttt{c1} & 1500 m/s & 1400--1600 m/s \\
\texttt{beta1} & 100 & 50--150 \\
\texttt{A1} & 1.0 & 0.5--1.5 \\

\texttt{r2} & 0.03 m & 0.02--0.05 m \\
\texttt{c2} & 1540 m/s & 1500--1600 m/s \\
\texttt{beta2} & 300 & 200--400 \\
\texttt{A2} & 0.8 & 0.5--1.0 \\

\midrule
\multicolumn{3}{l}{\textit{Additive Noise (SNR Control)}} \\
\midrule
\texttt{snr\_db} & 10 dB & 5--20 dB \\

\midrule
\multicolumn{3}{l}{\textit{Fetal Movement Artifacts}} \\
\midrule
\texttt{movement\_enabled} & true & true/false \\
\texttt{movement\_intensity} & 1.3 & 1.0--2.0 \\
\texttt{movement\_rate\_per\_min} & 8.0 & 5--15 \\
\texttt{movement\_duration\_range} & (0.12, 0.45) s & (0.1, 0.5) s \\
\texttt{movement\_band} & (15, 200) Hz & (10, 300) Hz \\
\texttt{movement\_thump\_prob} & 0.35 & 0.2--0.5 \\

\midrule
\multicolumn{3}{l}{\textit{Uterine Contractions}} \\
\midrule
\texttt{uc\_enabled} & true & true/false \\
\texttt{uc\_rate\_per\_10min} & 4.0 & 2--6 \\
\texttt{uc\_duration\_range} & (10.0, 25.0) s & (5.0, 30.0) s \\
\texttt{uc\_rise\_fall\_frac} & (0.35, 0.35) & (0.3, 0.4) \\
\texttt{uc\_attenuation} & 0.45 & 0.3--0.6 \\
\texttt{uc\_noise\_band} & (0.5, 18.0) Hz & (0.5, 20.0) Hz \\
\texttt{uc\_noise\_intensity} & 0.8 & 0.5--1.0 \\

\bottomrule
\end{tabular}
\caption{INI-configurable parameters, default values, and suggested value ranges of the simulator.}
\label{tab:user_params_default}
\end{table}



\bibliographystyle{IEEEtran}
\bibliography{bibliography2}

@article{ForemanMackey2013emcee,
  author  = {Foreman-Mackey, Daniel and Hogg, David W. and Lang, Dustin and Goodman, Jonathan},
  title   = {emcee: The {MCMC} Hammer},
  journal = {Publications of the Astronomical Society of the Pacific},
  year    = {2013},
  volume  = {125},
  number  = {925},
  pages   = {306--312},
  doi     = {10.1086/670067}
}

@article{GoodmanWeare2010Ensemble,
  author  = {J. Goodman and J. Weare},
  title   = {Ensemble samplers with affine invariance},
  journal = {Communications in Applied Mathematics and Computational Science},
  volume  = {5},
  number  = {1},
  pages   = {65--80},
  year    = {2010},
  doi     = {10.2140/camcos.2010.5.65}
}

@INPROCEEDINGS{Kovacs2005LongTerm,
  author={Kovacs, F. and Horvath, Cs. and Torok, M. and Hosszu, G.},
  booktitle={2005 IEEE Engineering in Medicine and Biology 27th Annual Conference}, 
  title={Long-term Phonocardiographic Fetal Home Monitoring for Telemedicine Systems}, 
  year={2005},
  volume={},
  number={},
  pages={3946-3949},
  doi={10.1109/IEMBS.2005.1615325}
}

@article{CESARELLI2012513,
title = {Simulation of foetal phonocardiographic recordings for testing of FHR extraction algorithms},
journal = {Computer Methods and Programs in Biomedicine},
volume = {107},
number = {3},
pages = {513-523},
year = {2012},
issn = {0169-2607},
doi = {https://doi.org/10.1016/j.cmpb.2011.11.008},
author = {M. Cesarelli and M. Ruffo and M. Romano and P. Bifulco},
}

@inproceedings{Chourasia2009DevelopmentOA,
  title     = {Development of a Signal Simulation Module for Testing of Phonocardiography Based Prenatal Monitoring Systems},
  author    = {Chourasia, Vijay S. and Tiwari, Anil Kumar},
  booktitle = {2009 Annual IEEE India Conference},
  year      = {2009},
  pages     = {1--4}
}

@article{Gerhardt2000FetalExposures,
  author  = {Gerhardt, Keith J. and Abrams, Robert M.},
  title   = {Fetal exposures to sound and vibroacoustic stimulation},
  journal = {Journal of Perinatology},
  volume  = {20},
  pages   = {S21--S30},
  month   = dec,
  year    = {2000}
}

@inproceedings{Samieinasab2015BSS,
  author    = {Samieinasab, M. and Sameni, R.},
  title     = {Fetal phonocardiogram extraction using single channel blind source separation},
  booktitle = {2015 23rd Iranian Conference on Electrical Engineering},
  year      = {2015},
  pages     = {78--83}
}

@article{Bhaskaran2022HR,
  author  = {Bhaskaran, A. and J, S. K. and George, S. and Arora, M.},
  title   = {Heart rate estimation and validation algorithm for fetal phonocardiography},
  journal = {Physiological Measurement},
  volume  = {43},
  pages   = {075008},
  month   = jul,
  year    = {2022}
}

@article{moya2025challenges,
  title={Challenges and Advances in Digital Processing of Fetal Phonocardiography Signal: A Review},
  author={Moya-Albor, Ernesto and Brieva, Jorge and Gomez-Coronel, Sandra L and Renza, Diego},
  journal={Machine Learning Methods in Biomedical Field: Computer-Aided Diagnostics, Healthcare and Biology Applications},
  pages={161--187},
  year={2025},
  publisher={Springer}
}

@inproceedings{Ruffo2010ITAB,
  author    = {Ruffo, M. and Cesarelli, M. and Romano, M. and Bifulco, P. and Fratini, A.},
  title     = {A simulating software of fetal phonocardiographic signals},
  booktitle = {Proc. 10th IEEE Int. Conf. on Information Technology and Applications in Biomedicine (ITAB)},
  year      = {2010},
  pages     = {1--4},
  doi       = {10.1109/ITAB.2010.5687716}
}

@article{ruffo2010algorithm,
  title={An algorithm for FHR estimation from foetal phonocardiographic signals},
  author={Ruffo, Mariano and Cesarelli, Mario and Romano, Maria and Bifulco, Paolo and Fratini, Antonio},
  journal={Biomedical Signal Processing and Control},
  volume={5},
  number={2},
  pages={131--141},
  year={2010},
  publisher={Elsevier}
}

@article{Romano2015Tool,
  author  = {Romano, M. and Bifulco, P. and Iuppariello, L. and Clemente, F. and D'Addio, G. and Cesarelli, M.},
  title   = {A new tool for foetal phonocardiography simulation},
  journal = {Studies in Health Technology and Informatics},
  volume  = {210},
  year    = {2015},
  pages   = {743--747},
  note    = {PMID: 25991252}
}

@INPROCEEDINGS{Amrutha2019FoetalAcousticSimulator,
  author    = {Amrutha, B. and Sinha, Raveesh and Kumar, Sidhesh and Arora, Manish},
  booktitle = {2019 11th International Conference on Communication Systems \& Networks (COMSNETS)},
  title     = {Foetal Acoustic Simulator},
  year      = {2019},
  pages     = {882--885},
  doi       = {10.1109/COMSNETS.2019.8711398}
}

@article{von2013normal,
  title={What is the “normal” fetal heart rate?},
  author={Von Steinburg, Stephanie Pildner and Boulesteix, Anne-Laure and Lederer, Christian and Grunow, Stefani and Schiermeier, Sven and Hatzmann, Wolfgang and Schneider, Karl-Theodor M and Daumer, Martin},
  journal={PeerJ},
  volume={1},
  pages={e82},
  year={2013},
  publisher={PeerJ Inc.}
}

@article{ramadan2020characteristics,
  title={Characteristics of fetal and maternal heart rate tracings during labor: A prospective observational study},
  author={Ramadan, Mohamad K and Fasih, Rana and Itani, Saadeddine and Salem Wehbe, Georges R and Badr, Dominique A},
  journal={Journal of neonatal-perinatal medicine},
  volume={12},
  number={4},
  pages={405--410},
  year={2020},
  publisher={SAGE Publications Sage UK: London, England}
}

@incollection{crellin2025sonography,
  title={Sonography evaluation of amniotic fluid},
  author={Crellin, Holly B and Singh, Vikramjeet},
  booktitle={StatPearls [Internet]},
  year={2025},
  publisher={StatPearls Publishing}
}

@article{alalaf2022intrapartum,
  title={Intrapartum ultrasound measurement of the lower uterine segment thickness in parturients with previous scar in labor: a cross-sectional study},
  author={Alalaf, Shahla K and Mansour, Tarek Mohamed M and Sileem, Sileem Ahmad and Shabila, Nazar P},
  journal={BMC Pregnancy and Childbirth},
  volume={22},
  number={1},
  pages={409},
  year={2022},
  publisher={Springer}
}

@article{lai2016fetal,
  title={Fetal movements as a predictor of health},
  author={Lai, Jonathan and Nowlan, Niamh C and Vaidyanathan, Ravi and Shaw, Caroline J and Lees, Christoph C},
  journal={Acta obstetricia et gynecologica Scandinavica},
  volume={95},
  number={9},
  pages={968--975},
  year={2016},
  publisher={Wiley Online Library}
}

@inproceedings{liu2010uterine,
  title={Uterine contraction modeling and simulation},
  author={Liu, Miao and Belfore, Lee A and Shen, Yuzhong and Scerbo, Mark W},
  booktitle={Selected Papers Presented at MODSIM World 2009 Conference and Expo},
  year={2010}
}

@article{goldberger2000physiobank,
  title={PhysioBank, PhysioToolkit, and PhysioNet: components of a new research resource for complex physiologic signals},
  author={Goldberger, Ary L and Amaral, Luis AN and Glass, Leon and Hausdorff, Jeffrey M and Ivanov, Plamen Ch and Mark, Roger G and Mietus, Joseph E and Moody, George B and Peng, Chung-Kang and Stanley, H Eugene},
  journal={circulation},
  volume={101},
  number={23},
  pages={e215--e220},
  year={2000},
  publisher={Lippincott Williams \& Wilkins}
}

@article{Kovacs2011Review,
  author  = {Kov{\'a}cs, Ferenc and Horv{\'a}th, Csaba and Balogh, {\'A}d{\'a}m T. and Hossz{\'u}, G{\'a}bor},
  title   = {Fetal phonocardiography---past and future possibilities},
  journal = {Computer Methods and Programs in Biomedicine},
  volume  = {104},
  number  = {1},
  pages   = {19--25},
  year    = {2011}
}

@article{adithya2017trends,
  title={Trends in fetal monitoring through phonocardiography: Challenges and future directions},
  author={Adithya, Prashanth Chetlur and Sankar, Ravi and Moreno, Wilfrido Alejandro and Hart, Stuart},
  journal={Biomedical Signal Processing and Control},
  volume={33},
  pages={289--305},
  year={2017},
  publisher={Elsevier}
}

@article{welch2003use,
  title={The use of fast Fourier transform for the estimation of power spectra: A method based on time averaging over short, modified periodograms},
  author={Welch, Peter},
  journal={IEEE Transactions on audio and electroacoustics},
  volume={15},
  number={2},
  pages={70--73},
  year={2003},
  publisher={IEEE}
}

@article{Barnova2022Comparative,
  author  = {Barnova, Katerina and Kahankova, Radana and Jaros, Rene and Litschmannova, Martina and Martinek, Radek},
  title   = {A comparative study of single-channel signal processing methods in fetal phonocardiography},
  journal = {PLOS ONE},
  volume  = {17},
  number  = {8},
  pages   = {e0269884},
  year    = {2022},
  month   = aug,
  doi     = {10.1371/journal.pone.0269884}
}

@article{Kahankova2023Review,
  author  = {Kahankova, Radana and Mikolasova, Martina and Jaros, Rene and Barnova, Katerina and Ladrova, Martina and Martinek, Radek},
  title   = {A Review of Recent Advances and Future Developments in Fetal Phonocardiography},
  journal = {IEEE Reviews in Biomedical Engineering},
  volume  = {16},
  pages   = {653--671},
  year    = {2023}
}

@article{muller2025standardized,
  title={Standardized Evaluation of Fetal Phonocardiography Processing Methods},
  author={M{\"u}ller, Krist{\'o}f and Hatvani, Janka and Goda, M{\'a}rton {\'A}ron and Koller, Mikl{\'o}s},
  journal={arXiv preprint arXiv:2507.10783},
  year={2025}
}

@article{Virtanen2020SciPy,
  author  = {Virtanen, Pauli and Gommers, Ralf and Oliphant, Travis E. and others},
  title   = {SciPy 1.0: fundamental algorithms for scientific computing in Python},
  journal = {Nature Methods},
  volume  = {17},
  pages   = {261--272},
  year    = {2020},
  doi     = {10.1038/s41592-019-0686-2}
}

@inproceedings{Choklati2022LaplacePCG,
  author    = {Abdelouahad Choklati and Anas Had and Khalid Sabri},
  title     = {On the Modelling of Phonocardiogram Signals: {L}aplace Kernel and Cyclostationarity Based Approaches},
  booktitle = {Nonstationary Systems: Theory and Applications},
  editor    = {Fakher Chaari and Jacek Leskow and Agnieszka Wylomanska and Radoslaw Zimroz and Antonio Napolitano},
  publisher = {Springer International Publishing},
  address   = {Cham},
  year      = {2022},
  pages     = {193--206},
  isbn      = {978-3-030-82110-4},
  doi       = {10.1007/978-3-030-82110-4_10}
}

@INPROCEEDINGS{Ali2011Dynamic,
  author={Almasi, Ali and Shamsollahi, Mohammad B. and Senhadji, Lotfi},
  booktitle={2011 Annual International Conference of the IEEE Engineering in Medicine and Biology Society}, 
  title={A dynamical model for generating synthetic Phonocardiogram signals}, 
  year={2011},
  volume={},
  number={},
  pages={5686-5689},
  doi={10.1109/IEMBS.2011.6091376}}


\end{document}